\documentclass{SBCbookchapter}
\usepackage{hanging}
\usepackage{authblk}
\usepackage{booktabs}
\usepackage{array}
\usepackage[utf8]{inputenc}
\usepackage[T1]{fontenc}
\usepackage[brazil,english]{babel}
\usepackage{graphicx}
\usepackage{booktabs}
\usepackage{fancyhdr}
\usepackage{lscape}
\usepackage{url}
 \usepackage{booktabs}
 \usepackage[normalem]{ulem}
 \useunder{\uline}{\ul}{}
 \usepackage{lscape}
 \usepackage{longtable}
 \usepackage[longnamesfirst, authoryear]{natbib}
 \usepackage{hyperref}
\usepackage{academicons}
\usepackage{xcolor}

\newcommand{\orcid}[1]{\href{https://orcid.org/#1}{\textcolor[HTML]{A6CE39}{\aiOrcid}}}

\author[1.2]{Xiaoming Zhai}  

\affil[1]{AI4STEM Education Center, University of Georgia}
\affil[2]{Department of Mathematics, Science, and Social Studies Education, University of Georgia}
\title{AI and Machine Learning for Next Generation Science Assessments}
\begin{document}
\maketitle  
\begin{abstract}

This chapter focuses on the transformative role of Artificial Intelligence (AI) and Machine Learning (ML) in science assessments. The paper begins with a discussion of the \textit{Framework for K-12 Science Education}, which calls for a shift from conceptual learning to knowledge-in-use. This shift necessitates the development of new types of assessments that align with the \textit{Framework's} three dimensions: science and engineering practices, disciplinary core ideas, and crosscutting concepts. The paper further highlights the limitations of traditional assessment methods like multiple-choice questions, which often fail to capture the complexities of scientific thinking and three-dimensional learning in science. It emphasizes the need for performance-based assessments that require students to engage in scientific practices like modeling, explanation, and argumentation. The paper achieves three major goals: reviewing the current state of ML-based assessments in science education, introducing a framework for scoring accuracy in ML-based automatic assessments, and discussing future directions and challenges. It delves into the evolution of ML-based automatic scoring systems, discussing various types of ML like supervised, unsupervised, and semi-supervised learning. These systems can provide timely and objective feedback, thus alleviating the burden on teachers.  The paper concludes by exploring pre-trained models like BERT and finetuned ChatGPT, which have shown promise in assessing students' written responses effectively.

\end{abstract}

% Corresponding author: Xiaoming Zhai \\
% Address: 125M Aderhold Hall, 110 Carlton Street  \\
% Athens, Georgia 30602
% Office: 517-432-0816 \\ 
% Email: Xiaoming.Zhai@uga.edu \\ \\
% \textbf{ORCID:} Zhai: https://orcid.org/0000-0003-4519-1931 \\ \\ 
% \textbf{Competing interests} \\
% The author declares no conflict of interest. \\ \\
% \textbf{Funding} \\
% This material is based upon work supported by the National Science Foundation (NSF) under Grant No. 2101104, 2138854. Any opinions, findings, and conclusions or recommendations expressed in this material are those of the author(s) and do not necessarily reflect the views of the NSF.

\begin{Cite This Chapter}
\begin{otherlanguage}{brazil}
Zhai, X. (2024). AI and Machine Learning for Next Generation Science Assessments. Jiao, H., \& Lissitz, R. W. (Eds.). Machine learning, natural language processing and psychometrics. Charlotte, NC: Information Age Publisher.

\end{otherlanguage}
\end{Cite This Chapter}

\newpage
\section{Introduction}

The rapid advancement of artificial intelligence (AI) in recent years has brought about transformative changes in various domains, including science assessment. Conventional methods of assessment in science education, particularly in classroom settings, often rely on multiple-choice questions (Zhai \& Li, 2021), which may not fully capture students’ understanding and engagement with scientific practices. This problem is even pronounced with the realization of three-dimensional learning in science—integrating science and engineering practices, disciplinary core ideas, and crosscutting concepts—a new vision set forth in the Framework for K-12 Science Education (National Research Council, 2012). This vision presents challenges for Next Generation Science Assessments because traditional assessments often fall short of capturing the complexities of scientific thinking during science and engineering practices. However, with the advent of machine learning (ML) techniques—an advanced artificial intelligence (AI), there is a growing opportunity to revolutionize the assessment practices in the science learning (Zhai, Haudek, Shi, et al., 2020). 

To assess Next Generation Science Learning and foster critical thinking and problem-solving skills, efforts have been put into developing ML- and performance-based assessments, which can engage students in science and engineering practices. Performance-based assessments usually require students to observe phenomena and develop explanations, arguments, or solutions (Harris et al., 2019). These assessment tasks require students to represent their thinking using multimodalities, such as writing or drawing (Zhai \& Nehm, 2023). However, writing and drawing are challenging to score in a timely fashion. For classroom assessment practices, without timely feedback, the promise of such assessments might be significantly compromised. That is, one of the primary reasons that teachers use assessments is to solicit information about students’ learning and, based on which, adjust instruction and make instructional decisions. It can be expected that teachers might be reluctant to use these complex assessment tasks if scores are not available in a timely fashion. In this case, technologies are desired. ML, due to its ability to automatically score written responses and drawn models, is promising, with which vast efforts in assessment development thus can be beneficial to millions of students (Jiao \& Lissitz, 2020; Zhai, Krajcik, et al., 2021). Moreover, with the increasing adoption of online learning platforms and the availability of vast amounts of educational data, the changing landscape of education creates an opportune moment for leveraging ML algorithms to improve science assessment practices (Linn et al., 2023; Zhai et al., 2023). By harnessing the power of ML, we can develop more personalized and adaptive assessments that provide students with timely feedback and promote deeper learning experiences.

This paper aims to explore the potential of ML-based assessments in shaping the next generation science education. This paper has three major goals. Firstly, we aim to review the evolution and current state of ML-based assessments in science education, identifying the challenges and opportunities they present. To achieve this, we will synthesize the existing research and highlight the various ML approaches and techniques used in this field. Secondly, we intend to introduce a framework accounting for ML-based assessments scoring accuracy in science learning environments. This framework will provide guidelines for educators and researchers to develop effective and accurate assessment tools that integrate ML algorithms. At last, we will discuss the most pressing issues of ML-based assessments in science education and the future directions. By achieving these goals, we hope to contribute to the ongoing efforts in advancing science education and paving the way for the ML-based Next Generation Science Assessments that better align with the vision of science learning set forth in the Framework.

%\begin{enumerate}
    %\item What AI Applications have been developed in the last 15 years to support students with learning disabilities?
   % \item How have these AI technologies been integrated into supporting students with learning disabilities in the classroom?\end{enumerate}
 
\section{Next Generation Science Assessments: A Shift from Conceptual Change to Knowledge-in-Use}

The Framework for K-12 Science Education (NRC, 2012) introduced significant reforms in science education by outlining a set of science and engineering practices, disciplinary core ideas, and crosscutting concepts to guide science instruction in K-12 classrooms. These reforms represent a major shift away from conceptual change to knowledge-in-use learning, advertising that scientific knowledge is meaningful for students only if students can apply the knowledge in designing solutions, constructing explanations, and solving problems. Rote memorization of science concepts or principles contributes limited to students’ science learning and the consequential success in their lives and careers. Therefore, Next Generation Science Learning should aim to enhance students’ scientific competencies and foster deeper thinking through practice-based learning. 

Achieving Next Generation Science Learning presents several challenges for classroom assessment practices. According to students’ grade levels and learning progression, the Next Generation Science Standards ([NGSS]; NGSS Lead States, 2013) specify a series of performance expectations, which provide benchmarks for learning. These performance expectations are three-dimensional, integrating science and engineering practices, crosscutting concepts, and disciplinary core ideas. Assessing such performance expectations thus becomes critical to facilitate classroom learning. However, traditional forms of assessments, such as multiple-choice exams, are no longer sufficient to evaluate students’ uses of scientific knowledge and their ability to engage in scientific practices. The NGSS call for more authentic assessments that reflect real-world scientific inquiry and the application of scientific knowledge. This shift towards authentic assessment poses challenges in terms of designing and administering assessments that accurately measure students’ abilities-- thinking critically, solving problems, and engaging in scientific practices. 

Assessments that need to align with the three dimensions outlined in the Framework and the performance expectations in the NGSS require the development of performance-based tasks, open-ended questions, and hands-on experiments. Researchers, such as Harris et al. (2019), developed evidence-centered approaches to crafting assessment tasks that are usually performance-based constructed responses. To assess students’ three-dimensional performance, they designed two- or three-level analytic rubrics specifically tailored to each item. These rubrics were created based on evidence-based guiding principles, aiming to evaluate student performance at the beginning, developing, and proficient levels. The determination of the number of rubric levels was influenced by the complexity of the aspects being evaluated.

To illustrate this challenge, we cite an example assessment task (i.e., Red dye diffusion) developed to assess Next Generation Science Learning (see Figure 1). This task targets one performance expectation from the NGSS (NGSS Lead States, 2013) at the middle school level: MS-PS1-4. Develop a model that predicts and describes changes in particle motion, temperature, and state of a pure substance when thermal energy is added or removed. The performance expectation requires students to be able to develop models that can predict and describe changes in particle motion, temperature, and state of a pure substance when thermal energy is added or removed. This task thus focuses on red dye particles’ motion in the water. To engage students in scientific practice, this task first provides students with a video that demonstrates dye diffusion in water at three different temperatures. Students are then asked to construct models to explain their observations from the video and provide a written description of their models. A drawing pad with various drawing tools, three drawing boxes, and a text box for writing is provided to facilitate this task (see Figure 1 for the task, answer space, and student example responses). 

\begin{figure}[htp]
\centering
\includegraphics[width=13cm]{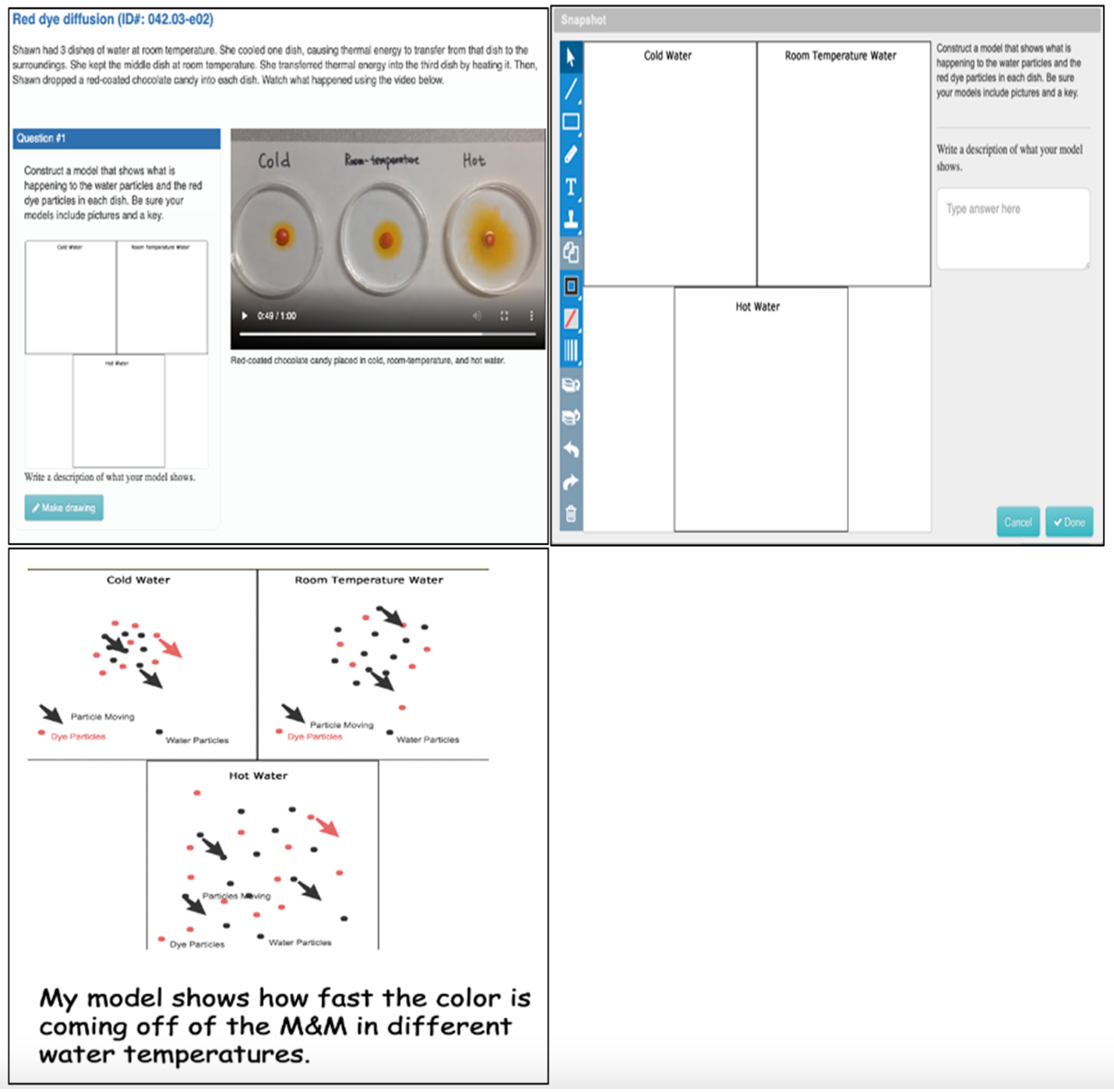}
\caption{Assessment task: “Red dye diffusion” item screenshot (left), response interface (right), and a student response (bottom)(adapted from (Zhai, He, \& Krajcik, 2022)) }
% \label{fig:galaxy}

\end{figure}

Performance-based assessments, such as Red dye diffusion, are critical to assess and facilitate Next Generation Science Learning, but place additional burdens on teachers to score these assessments. Specifically, teachers are not able to assign scores, evaluations, or feedback to every student in ongoing classrooms. Even for after-class use of the assessments, it is costly and time-consuming for teachers to score student responses. For example, assigning scores to each student’s drawn and written responses to the example item may averagely cost 1 minute. For a normal middle school science teacher who teaches 100 students (four classes and each with 25 students), it would cost 100 minutes to score this individual assessment task, which can be a factor in decreasing teachers’ motivation to use the assessment tasks. 

In summary, Next Generation Science Assessment practice is challenging not only in development and design but also scoring and use of the scores. The Framework brought about a shift in classroom assessment practices, necessitating authentic and automatically scored assessments aligned with the Framework’s three dimensions. New technologies that may enable the automatic scoring of complex assessments are desired.

%Learning disabilities, also known as neurodevelopmental disorders, are due to genetic or neurobiological factors that alter brain functions; thus, learning disabilities do not include any learning problems that may be due to visual, hearing, emotional, or motor disabilities, and it does not include any learning problems that may be due to environmental, cultural, or economic disadvantages (Learning Disabilities Association of America, n.d.; Individuals with Disabilities Education Act, 2007). The Individuals with Disabilities Education Act, a crucial piece of legislation in the United States, provides a specific definition for learning disability, 
%\begin{itemize}
    %\emph{A disorder in one or more of the basic psychological processes involved in understanding or in using language, spoken or written, that may manifest itself in the imperfect ability to listen, think, speak, read, write, spell, or to do mathematical calculations, including conditions such as perceptual disabilities, brain injury, minimal brain dysfunction, dyslexia, and developmental aphasia.} (Individuals with Disabilities Education Act, 2007, para. 10) 
%\end{itemize}

\section{Evolution of Machine Learning-based Automatic Scoring for Science Assessments}

Machine learning can address these challenges by entailing automatic scoring systems and providing timely and objective feedback to both students and teachers. ML algorithms can analyze students’ responses and performance on various assessment tasks, including written responses, lab reports, or project-based assignments. These algorithms can assist teachers in handling the time-consuming task of grading, assess the quality of students’ performance and thinking, and provide immediate feedback on their strengths and areas for improvement. These systems can provide consistent and efficient scoring, reducing teachers’ workload and allowing them to focus on providing personalized instruction and support.

The realm of science assessments has witnessed the application of five distinct stages of ML applied in science assessments: supervised, unsupervised, and semi-supervised ML, pre-trained model, and zero-short. All five types share a fundamental characteristic, which is the machine’s capacity to “learn” from accumulated “experience” rather than adhering to explicit human instructions (Zhai, Yin, et al., 2020). ML-based assessment applications generally encompass two stages: training/learning and testing/predicting. Supervised, unsupervised, semi-supervised ML, pre-trained model, and zero-short differ from one another based on their respective approaches to training. In the case of supervised ML, the training data are initially labeled by human annotators, and the machine learns from these labeled instances (examples see Haudek et al., 2012; Nehm et al., 2012). Supervised ML is the most frequently used approach in automatic scoring given its high accuracy. For example, our research has assessed student performance when engaging in scientific argumentation (Wilson et al., 2023; Zhai, Haudek, et al., 2022), investigation (Maestrales et al., 2021), explanations (Zhai, He, et al., 2022), etc., and achieved satisfactory scoring accuracy using supervised ML (see Figure 2). Our recent work employed the deep learning algorithms (e.g., neural networks) further extend the usability of this approach from language models to computer visions, and thus ML can automatically score student-drawn models (Wang et al., In press; Zhai, He, et al., 2022). In addition, my colleagues and I have applied supervised ML to assess teachers’ pedagogical content knowledge (Zhai, Haudek, Stuhlsatz, et al., 2020).

\begin{figure}[htp]
\centering
\includegraphics[width=13cm]{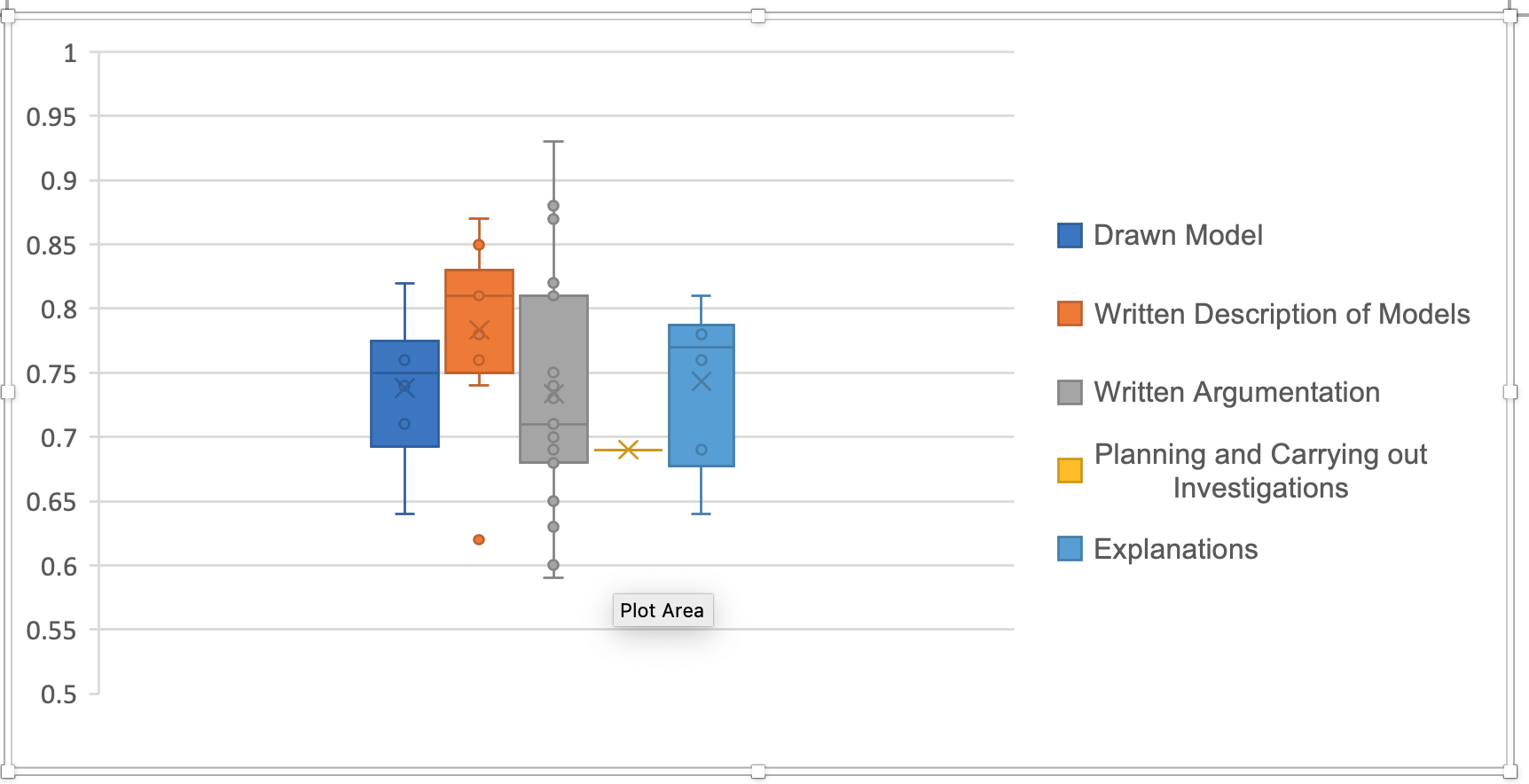}
\caption{ Automatic scoring accuracy for science assessments involving different scientific practices }
% \label{fig:galaxy}

\end{figure}

Conversely, unsupervised ML relies on unlabeled data, requiring the machine to detect underlying patterns by analyzing the structural properties within the data corpus. Although unsupervised ML has the potential to reduce resource consumption and costs by eliminating the need for human labeling, it may lead to decreased accuracy in scoring students’ responses, which can prove problematic for most forms of science assessments. In the earlier years, Urban-Lurain et al. (2013) have used unsupervised ML to assist in scoring rubric development, using the identified patterns from student responses. However, unsupervised ML is rarely used for scoring purposes by itself. Hybrid approaches, such as semi-supervised ML, have been developed to address challenges that need a large amount of labeled data, wherein humans employ a portion of labeled data to train the machine while also allowing it to learn the data structure from the unlabeled portions. Irrespective of the specific ML type employed, the development and validation of rigorous algorithms are standard practices before the algorithms can be deployed to predict/classify instances within new corpora.

Figure 3 provides a comprehensive outline of the general procedure employed for ML applications in science assessment. Commencing with science performance expectations, assessment tasks and rubrics are constructed by item developers. Subsequently, student responses to these tasks are collected. In cases where supervised or semi-supervised ML is utilized, human experts are enlisted to score student responses based on the provided rubrics (Unsupervised ML usually omits the human scoring part). All information derived from student responses, including human scores, is then utilized to train the machine. The training process involves extracting a set of features or attributes from student responses and establishing their association with the provided labels through a specific algorithmic model chosen by researchers. Researchers have the flexibility to fine-tune the parameters of the algorithmic model to align with their research objectives. The outcome of the training process is an optimized algorithmic model that can be employed to identify patterns within new datasets or categorize responses accordingly. However, prior to making predictions on new corpora, it is customary to undertake a model validation procedure. Common validation methods include self-validation, data splitting, or cross-validation.

Among the various validation approaches employed in science assessment, cross-validation is the most prevalent (Zhai, Yin, et al., 2020). Cross-validation entails partitioning the data into n groups and employing (n-1) groups for training the machine, while the remaining group is utilized to evaluate the algorithm’s performance. Specifically, the algorithm is employed to predict labels for the remaining group, and a comparison is made between the machine-assigned labels and the human-assigned labels to calculate a mean machine-human agreement (MHA). This process is repeated n times to ensure that each group has an opportunity to serve as the testing set. The average of the n MHAs serves as an indicator of the algorithm’s capability. If the average MHA satisfies predefined success criteria, the algorithm can be applied to new datasets. 

\begin{figure}[htp]
\centering
\includegraphics[width=13cm]{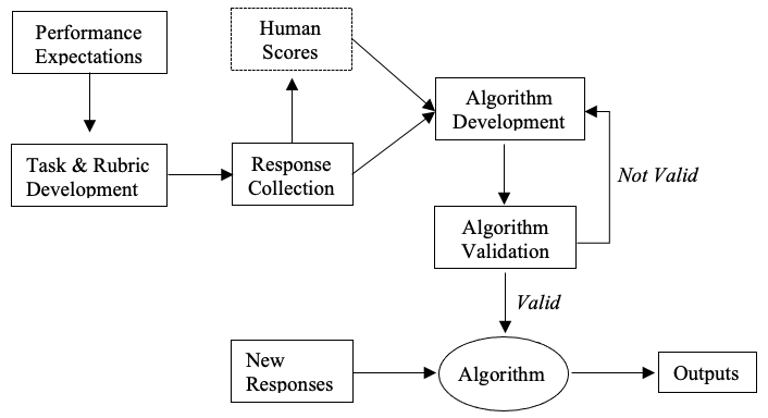}
\caption{Machine Learning-based Assessment Framework }
% \label{fig:galaxy}

\end{figure}

Pre-trained models in ML have emerged as powerful tools that offer significant advantages in various domains. These models are pre-trained on vast amounts of high-quality data and possess the ability to learn intricate patterns and representations. By leveraging the knowledge and expertise acquired during pre-training, these models can be fine-tuned or utilized as feature extractors for specific tasks, leading to improved performance and reduced training time. One prominent example of a pre-trained model is BERT (Bidirectional Encoder Representations from Transformers). BERT is a language model developed by Google that has achieved remarkable success in various natural language processing (NLP) tasks (Devlin et al., 2018). It is pre-trained on a large corpus of text data, such as Wikipedia, which allows it to capture the contextual relationships between words and understand the meaning of sentences. Riordan et al. (2020) found that BERT can rival the performance of traditional ML algorithms (e.g., SVR, Neural Networks) in assessing students’ written responses to science assessments. Once pre-trained, BERT can be fine-tuned on specific downstream tasks, such as text classification, named entity recognition, or sentiment analysis. During fine-tuning, BERT’s parameters are adjusted using task-specific data, enabling the model to specialize in a particular NLP task. Liu et al. (2023) used more than 5,0000 student-written responses to fine-tune (unsupervised ML) a SciEdBERT and found that this fine-tuned BERT performed better than the general BERT in terms of developing scoring models. This is because, one of the key advantages of BERT is its ability to comprehend the context and meaning of words based on the surrounding words in a sentence, considering both the left and right contexts. This bidirectional understanding makes BERT particularly effective in tasks that require a deep understanding of language nuances. Liu et al. (2023) took advantage of BERT as a pre-trained model and further enhanced its ability to comprehend student writing in science education. Developers and researchers can benefit from its extensive language understanding capabilities without the need to train a model (supervised ML) from scratch. This not only saves significant computational resources and time but also allows for faster development and deployment of NLP applications across a wide range of domains, including information retrieval, question answering, and language translation. Their versatility and effectiveness make pre-trained models a valuable asset in accelerating and enhancing ML applications across a wide range of fields.

Zero-shot ML refers to a paradigm where a model is able to perform a task or make predictions on classes or concepts that are not seen during training. It involves transferring knowledge from seen classes to unseen classes by leveraging semantic relationships or embeddings. In zero-shot learning, the model is trained on a labeled dataset with a subset of classes, referred to as “seen classes.” The model learns to understand the relationship between the input features and the class labels. During training, the model also learns to associate auxiliary information or semantic embeddings with each class, which captures the underlying semantic relationships between classes. Once trained, the zero-shot learning model can generalize to predict or classify instances from “unseen classes” that were not present in the training set. Wu et al. (2023) developed a zero-shot approach-- matching exemplars as next sentence prediction (MeNSP), using pre-trained models. Without training data, MeNSP was able to score student written responses with Cohen’s Kappa ranges between 0.30 to 0.57. This is achieved by utilizing the learned semantic embeddings or auxiliary information to reason about the unseen classes based on their relationships with the seen classes. With few-shot (i.e., a few training samples), MeNSP can increase the lower end of scoring accuracy of Cohen’s Kappa from 0.30 to 0.38. This approach significantly reduced the machine training efforts and may be used for low-stakes settings. 

More recently, the availability of fine-tuned ChatGPT shows substantial potential to improve scoring accuracy. In their research, Latif and Zhai (2013) investigates how a customized version of ChatGPT (GPT-3.5) can be effectively used for the automated grading of students' written answers in science education assessments. Although GPT-3.5 has shown remarkable capabilities in understanding natural language, its general training on a wide array of internet-based texts calls for additional fine-tuning to suit the specific language used by students in educational settings. To tackle this, they fine-tuned GPT-3.5 on six distinct assessment tasks, incorporating a variety of responses from middle and high school students, as well as expert evaluations. These tasks included both multi-label and multi-class question formats. They then compared the performance of their specialized GPT-3.5 with a fine-tuned version of Google's BERT model. Their results indicate that the fine-tuned GPT-3.5 surpassed BERT, achieving an average accuracy increase of 9.1\% across all tasks, and this improvement was statistically significant (p < 0.05). Particularly in multi-label tasks, GPT-3.5 demonstrated a 7.1\% higher accuracy rate than BERT for one specific item. In summary, their study confirms that a domain-specific GPT-3.5 model can effectively and accurately grade student responses in the realm of science education. To foster community engagement and further research, they have made these fine-tuned models publicly accessible.

The key distinction between zero-shot learning and pre-trained models lies in the nature of the transferred knowledge. Pre-trained models, such as those based on transfer learning, are initially trained on a large-scale dataset to learn general features and representations of the data. These models are then fine-tuned on a specific task using labeled or unlabeled data related to that task. The transferred knowledge in pre-trained models comes from the learned general representations. In contrast, zero-shot learning focuses on transferring knowledge about semantic relationships or embeddings between classes. The model learns to understand the underlying connections between classes during training, allowing it to generalize to unseen classes during inference. Both pre-trained models and zero-shot learning leverage prior knowledge, but they differ in the type of knowledge transferred. Pre-trained models transfer general feature representations, while zero-shot learning transfers knowledge about semantic relationships or embeddings. The choice between these approaches depends on the specific requirements and availability of labeled data for the task at hand.

Deploying ML and automatic scoring in the context of science assessments also presents its own set of challenges. One major challenge is ensuring the validity of automated scoring systems, particularly for complex and open-ended tasks (Zhai, Krajcik, et al., 2021). The algorithms need to be trained on large datasets of well-annotated responses to accurately assess students’ work and provide meaningful feedback. Additionally, concerns regarding the fairness and bias of automated scoring systems need to be addressed. ML algorithms should be developed and tested to ensure they are not biased against certain groups of students based on their race, gender, or socioeconomic status. To achieve this goal, it is essential to study the potential factors holding accountable for machine scoring accuracy.

\section{A Framework Accounting for Automatic Scoring Accuracy}

The most critical question for ML-based Next Generation Science Assessments is scoring accuracy, as it is the foundation to secure unbiased scores and fair uses of the scores. Machine scoring accuracy is usually measured by comparing machine scores with human consensus scores, indicated by Machine-Human Agreements (MHAs). In our prior work (Zhai, Shi, et al., 2021), we reviewed the literature and meta-analyzed the factors impacting machine scoring accuracy. In this study, we presented this extended framework that identifies five categories of factors moderating MHAs: (i) assessment external features, (ii) assessment internal features, (iii) examinee features (e.g., grade, school level, ELL), (iv) machine training and validation approaches (e.g., sample size, human rater reliability), and (v) technical features (e.g., algorithm, attribute abstraction).

\emph{Assessment external features}, such as length of responses, rubrics, assessment scenarios, and type of assessment, have been proposed as factors moderating MHAs. Length of student responses is regarded as a critical factor that limits MHAs because longer responses tend to provide more information for ML so that ML scoring may be more accurate. Therefore, it is not surprising that Nehm et al. (2012) found that scoring models were most effective in scoring longer responses, though longer responses are typically more challenging to score for human raters given their high cognitive demand for scoring. Despite this, it is noted that Nehm et al. (2012) only reported a weak relationship between response length and MHA. Scenarios or prompts are also external features that might moderate machine performance. Clear scoring criteria are necessary to achieve high MHAs (Lottridge et al., 2018). Many of the assessment external features have been examined in the science education literature, including the subject domain which was found to significantly impact MHAs (Zhai, Shi, et al., 2021). 

\emph{Assessment internal features}, such as the number of concepts, complexity of construct, degree of overlap of concepts, and variation in concepts, have been shown to moderate MHAs.. Some studies have explored the factors contributing to MHAs in text-based responses involving concepts. Williamson et al. (2010) proposed three key factors: the number of concepts elicited by an item, the variety of ways used to express these concepts, and the distinctness of the concept expressions. Leacock et al. (2013) suggested that a smaller number of well-defined concepts could contribute to MHAs, while Lottridge et al. (2018) identified the number, complexity, degree of overlap, and variation in concepts as moderators of MHAs. To specify the machine scoring ability, Lottridge et al. (2018) developed a comprehensive five-level framework for rating machine scoring ability. Item difficulty and complexity of constructs might also moderate MHAs, but few studies have provided empirical evidence to support this claim. McGraw-Hill Education (2014) discovered that machine performance decreased when scoring more difficult items, particularly those tapping into higher-order knowledge. Zhai, Haudek, Shi, et al. (2020) argued that the complexity of constructs may moderate MHAs, which was verified by Haudek and Zhai (2021) using empirical evidence. 

\emph{Examinee features}, such as grade, school level, and English language learners (ELLs), have been a topic of particular concern for machine scoring. However, not all studies have found that machine scoring is sensitive to examinee features. Examinee characteristics, including grade level and school level, have been the focus of machine scoring research in recent years (Zhai, Shi, et al., 2021). The impact of examinee diversity on machine scoring has been the subject of particular concern, with ELLs and lower-grade band students being especially vulnerable to scoring discrepancies (Wilson et al., 2023). Previous studies, such as Bridgeman et al. (2012), compared machine and human scores of Graduate Record Examinations (GRE) and the Test of English as a Foreign Language (TOEFL) assessments. These studies have reported varying scores for ELLs from different countries that were scored by machines and humans, with some languages being scored higher by machines than humans, and vice versa. 

While some studies, such as Liu et al. (2016), have found the difference in machine scores between ELLs and non-ELLs to be negligible, Ha and Nehm (2016a; 2016b) have found that ELLs’ constructed responses contain twice as many machine-scored words (MSW) than non-ELLs. However, MSW was found to be a relatively uncommon occurrence in student responses, comprising only 2\% of the corpus, and therefore, had minimal impact on machine-scored assessments. A recent study shows that machine scoring has the potential to enlarge the gap in average scores between ELLs and other students on science assessments (Wilson et al., 2023). Again, this effect only happened at a very small number of assessment items. For other examinee factors such as gender, Mao et al. (2018) have found that machine scoring is not as sensitive to factors such as gender differences compared to human raters.

\emph{Machine training and validation approaches} have been examined in previous studies and yielded inconsistent results. The techniques used to train and validate machine algorithms have been a subject of debate, with notable differences in the outcomes of different approaches. Specifically, Nehm et al. (2012) applied two training approaches to examine MHAs. With one set of items, they used the individual item responses to train the machine and then validated the algorithms using the same set of responses. In a second study, they used one set of responses to train the machine and then validated the algorithms using another set of responses. Although both approaches obtained satisfactory Cohen’s $\kappa$, they found a decrease in MHAs from the latter approach. This finding suggests that the choice of the training and validation approach could significantly affect the accuracy of MHAs. Nehm and Haertig (2012) also suggested potential relationships between machine and human rater performance on scoring based on item types. Lottridge et al. (2018) argued that rater behavior in labeling training data is a critical factor accounting for machine capability. However, tests of this assumption have yielded inconsistent results. For example, Bridgeman et al. (2012) found that MHAs can be high even though the machine scoring accuracy was lower than the human–human agreement.

Given that more studies have employed cross-validation approaches, rather than self- and split-validation, a prior review study (Zhai, Yin, et al., 2020) compared the MHAs generated using these different validation approaches. Cross-validation was found to yield higher MHAs compared to both self- and split-validation approaches. Training size of the corpus is another factor that has been widely discussed as a contributor to MHA (e.g., Nehm et al. 2012). One approach to account for these differences in training sample size is to consider a weighted mean Cohen’s $\kappa$. The use of a weighted mean Cohen’s $\kappa$ could address the challenges of varying training sample sizes and enhance the accuracy of MHAs.

\emph{Technical features}, such as algorithm types or attribute abstraction approaches, are also regarded as crucial factors contributing to diverse MHAs. Algorithmic models, for instance, have been found to be most critical to machine scoring accuracy (Zhai, Shi, et al., 2021), suggesting that the assessment used and algorithm models developed in the study may limit the generalizability of findings. Attributes are those features inherent to student responses that may be used to link student performance and scores. These attributes are the foundation of machine algorithm development. Lintean et al. (2012) compared two attribute abstraction approaches using different grain sizes and found that algorithms developed using individual words as attributes over-performed those developed using a set of words as a unit. Additionally, increasing the number of features was found to positively impact machine-based assessment performance. Although several technical features have been examined by Zhai, Shi et al. (2021), the most updated ML, such as pre-trained or zero-shot approaches – have not been thoroughly investigated. As such, little is currently known about how the most updated ML algorithms moderate machine-based assessment performance.

The above review of the literature has revealed considerable inconsistencies regarding how various factors moderate MHAs. This is because most studies only employed a limited number of assessment tasks, and the tests themselves were developed and validated using different approaches. Identifying variables that different studies have in common and are suitable for review would help to advance work in this growing area. Our meta-analysis has contributed to synthesizing available evidence and examining claims about the factors that contribute to MHAs in science assessment, but more research is needed to include the most updated ML algorithms and study the factors accounting for the variable MHAs in future work.

 \section{Issues and Future Directions }

While AI and ML have significantly extended the functionality of science assessment, been able to assess complex constructs, and ease humans’ efforts (Zhai, 2021), there are pressing issues both theoretically and practically that hinder the in-depth uses of ML-based science assessments. Addressing these issues is essential to move the field forward.

\subsection{Model Generalizability}

As ML techniques continue to advance, ensuring model generalizability becomes an important area of focus for Next Generation Science Assessments. In the current landscape, ML-based assessments largely rely on individual algorithmic models trained to assign scores for individual assessment tasks. Training individual algorithmic models is costly, and the problem needs to be addressed to make ML-based assessments available to a broader audience. While current models show promise in their ability to make accurate predictions on unseen data, there is still a need to improve their generalizability across different contexts, populations, and assessment tasks. Future research should explore strategies to enhance the transferability of ML models, such as incorporating more diverse and representative training data, leveraging domain adaptation techniques, and developing robust feature selection methods. Additionally, investigating the impact of model architecture and hyperparameter tuning on generalizability will be crucial in building more reliable and adaptable assessment models.

Moreover, advancements in federated learning can also contribute to model generalizability. Federated learning allows models to be trained collaboratively across multiple institutions or organizations without sharing raw data, thereby addressing issues related to data privacy and data distribution heterogeneity. Exploring federated learning approaches in the context of science assessments can lead to more robust models that can adapt to diverse educational settings while preserving data privacy.

\subsection{Unbalanced Data}

Addressing the issue of unbalanced data is another key direction for future advancements in ML-based science assessments. Many real-world assessment datasets suffer from class imbalance, where certain classes or categories of responses are significantly underrepresented (Jiao et al., 2023). This poses challenges for training accurate and unbiased models, as the learning algorithms may become biased towards the majority class without further examinations (Zhai \& Nehm, 2023).

To overcome this, researchers should explore techniques like oversampling minority classes, undersampling majority classes, or using advanced sampling methods like SMOTE (Synthetic Minority Over-sampling Technique). Fang, Lee, and Zhai (2013) employed GPT-4 to augment student written responses and found the GPT-4 augmented model performance was higher or equal to  the models trained using equal numbers of student written responses. The development of novel loss functions that explicitly account for class imbalance can improve the model’s ability to handle skewed datasets. Moreover, exploring ensemble learning methods that combine multiple models trained on balanced subsets of data can lead to improved performance and better handling of imbalanced classes.

Furthermore, as the field progresses, it will be essential to address intersectional biases that may arise from the combined effects of class imbalance and other demographic factors. Researchers should investigate techniques that consider the interplay between different demographic variables, such as gender, race, and socioeconomic status, to ensure that ML-based science assessments are fair and equitable for all students.

\subsection{Machine Learning-based Assessment User Guidelines }

As ML-based assessments become more prevalent, it is crucial to develop user guidelines and best practices to ensure their effective and ethical use. Educators, administrators, and policymakers need guidance on how to integrate these assessments into existing educational frameworks, interpret their results, and make informed decisions based on the outcomes (Zhai \& Krajcik, 2022). Future research should focus on creating comprehensive guidelines that address issues like test administration, data privacy, transparency in algorithmic decision-making, and potential biases and limitations of the models. These guidelines should also highlight the importance of human judgment and expertise in conjunction with ML-based assessments. Collaboration between researchers, educational practitioners, and policymakers will be vital in developing user-friendly tools and resources to support the successful implementation of ML-based science assessments in educational settings.

Additionally, as technology advances, it is crucial to continuously monitor and update guidelines to address emerging challenges and ethical considerations. For example, guidelines should adapt to advancements in explainable AI techniques, ensuring that users can understand how the ML models arrive at their predictions. The inclusion of guidelines for continuous model monitoring and evaluation will also enable educators and administrators to identify and address potential biases or limitations in real-time, fostering transparency and accountability in ML-based science assessments.

\subsection{Interpretations and Uses of Scores }

The interpretations and uses of scores obtained from ML-based science assessments require careful consideration to ensure their meaningful and responsible utilization in science education. As these assessments provide predictions or classifications based on complex algorithms, it is necessary to establish clear guidelines on how to interpret and communicate these scores effectively to stakeholders, such as students, teachers, parents, and policymakers. Future research should explore methods for generating interpretable explanations for the predictions made by ML models, allowing users to understand the underlying reasoning and factors contributing to the scores. Techniques such as attention mechanisms, saliency maps, and rule-based explanations can provide insights into the features or patterns that influenced the model’s decision, enhancing the transparency and trustworthiness of the assessment process.

Furthermore, investigating how these scores can be effectively integrated into educational decision-making processes, such as personalized instruction, curriculum design, and student support systems, will be essential for maximizing the benefits of ML-based science assessments. By aligning the assessment results with specific instructional interventions, educators can provide targeted support to students, identify areas for improvement, and foster individualized learning experiences. However, it is crucial to balance the use of ML-based scores with holistic assessments that consider multiple dimensions of student performance, including non-cognitive skills and qualitative feedback, to ensure a comprehensive understanding of student abilities (Zhai \& Nehm, 2023).

To support the responsible use of ML-based scores, the development of comprehensive data literacy programs for educators, students, and other stakeholders is necessary. These programs should focus on enhancing the understanding of the limitations and potential biases of ML models, promoting critical thinking about assessment results, and fostering informed decision-making based on multiple sources of evidence.

In conclusion, the future of ML for Next Generation Science Assessments lies in enhancing model generalizability, addressing unbalanced data challenges, developing user guidelines, and improving the interpretations and uses of scores. By tackling these key areas, researchers and practitioners can pave the way for more reliable, fair, and meaningful assessments that leverage the power of ML to enhance science education and foster deeper understanding and engagement among students.

.

%\begin{table}[h]
   % \centering
   % \caption{Inclusion and Exclusion Criteria}
   % \begin{tabular}{m{16em} m{16em}}
    %\hline
    %Inclusion Criteria & Exclusion Criteria \\
   % \hline \\
       %The source is a journal article or conference proceeding.   &  The source is something other than a journal article or conference proceeding. The source is a review study.   \\ 
       
       % Published in English  &  Not in the English language \\ [1ex]

       % The study involves artificial intelligence.  & The study does not involve artificial intelligence.  \\ [1ex]

       % The study targets students with learning disabilities, which includes dyslexia, dyscalculia, and dysgraphia.     &The study is targeting students without disabilities or with other disabilities. \\ [1ex]  

       % Content being looked at is reading, writing, or math in the English language.     &  Content being looked at is reading, writing, or math in a language other than English. \\ [1ex]

       % The study is related or conducted to the field of education or oriented towards supporting students' education. &The study is related to the field of medicine, taking place in a clinical or medical setting, or requiring medical equipment or personnel. \\ [1ex]
        
        %The study provides information on supporting, instructing, or assessing students with learning disabilities.  &   The study is focused solely on screening, diagnosing, identifying, predicting, or classifying a learning disability. \\ [1ex]        
       % \hline 
   % \end{tabular}
    
   % \label{tab:my_label}
%\end{table}

\

\section*{Acknowledgement}

This presentation was funded by National Science Foundation(NSF) (Award no. 2101104, 2138854). Any opinions, findings, conclusions, or recommendations expressed in this material are those of the author(s) and do not necessarily reflect the views of the NSF.

% \newpage
\section*{References }
\begin{hangparas}{.5in}{1}
Bridgeman, B., Trapani, C., \& Attali, Y. (2012). Comparison of human and machine scoring of essays: Differences by gender, ethnicity, and country. Applied Measurement in Education, 25(1), 27-40. 

Devlin, J., Chang, M.-W., Lee, K., \& Toutanova, K. (2018). Bert: Pre-training of deep bidirectional transformers for language understanding. arXiv preprint \url{arXiv:1810.04805}. 

Fang, L., Lee, G., \& Zhai, X. (2023). Using GPT-4 to Augment Unbalanced Data for Automatic Scoring. arXiv preprint arXiv:5194005 

Ha, M., \& Nehm, R. (2016a). Predicting the accuracy of computer scoring of text: Probabilistic, multi-model, and semantic similarity approaches. Paper in proceedings of the National Association for Research in Science Teaching, Baltimore, MD, April, 14-17. 

Ha, M., \& Nehm, R. H. (2016b). The impact of misspelled words on automated computer scoring: a case study of scientific explanations. Journal of Science Education and Technology, 25(3), 358-374.\url{https://link-springer-com.proxy1.cl.msu.edu/content/pdf/10.1007%2Fs10956-015-9598-9.pdf} 

Harris, C. J., Krajcik, J. S., Pellegrino, J. W., \& DeBarger, A. H. (2019). Designing knowledge-in-use assessments to promote deeper learning. Educational measurement: issues and practice, 38(2), 53-67. \url{https://doi.org/10.1111/emip.12253} 

Haudek, K. C., Prevost, L. B., Moscarella, R. A., Merrill, J., \& Urban-Lurain, M. (2012). What are they thinking? Automated analysis of student writing about acid–base chemistry in introductory biology. CBE—Life Sciences Education, 11(3), 283-293. 

Haudek, K. C., \& Zhai, X. (2021). Exploring the Effect of Assessment Construct Complexity on Machine Learning Scoring of Argumentation National Association of Research in Science Teaching, Florida. 

Jiao, H., \& Lissitz, R. (2020). Application of Artificial Intelligence to Assessment. Charlotte, NC: Information Age Publishing. 

Jiao, H., Yadav, C., \& Li, G. (2023). Integrating Psychometric Analysis and Machine Learning to Augment Data for Cheating Detection in Large-Scale Assessment. 

Latif, E., \& Zhai, X. (2023). Fine-tuning ChatGPT for Automatic Scoring. arXiv preprint arXiv:2310.10072.

Leacock, C., Messineo, D., \& Zhang, X. (2013). Issues in prompt selection for automated scoring of short answer questions. Annual conference of the National Council on Measurement in Education, San Francisco, CA, 

Linn, M. C., Donnelly-Hermosillo, D., \& Gerard, L. (2023). Synergies Between Learning Technologies and Learning Sciences: Promoting Equitable Secondary School Teaching. In Handbook of research on science education (pp. 447-498). Routledge. 

Lintean, M., Rus, V., \& Azevedo, R. (2012). Automatic detection of student mental models based on natural language student input during metacognitive skill training. International Journal of Artificial Intelligence in Education, 21(3), 169-190. 

Liu, O. L., Rios, J. A., Heilman, M., Gerard, L., \& Linn, M. C. (2016). Validation of automated scoring of science assessments. Journal of Research in Science Teaching, 53(2), 215-233. 

Liu, Z., He, X., Liu, L., Liu, T., \& Zhai, X. (2023). Context Matters: A Strategy to Pre-train Language Model for Science Education. In N. Wang \& e. al. (Eds.), AI in Education 2023 (Vol. CCIS 1831, pp. 1-9). Springer. \url{https://doi.org/10.1007/978-3-031-36336-8_103 }

Lottridge, S., Wood, S., \& Shaw, D. (2018). The effectiveness of machine score-ability ratings in predicting automated scoring performance. Applied Measurement in Education, 31(3), 215-232. 

Maestrales, S., Zhai, X., Touitou, I., Baker, Q., Krajcik, J., \& Schneider, B. (2021). Using machine learning to score multi-dimensional assessments of chemistry and physics. Journal of Science Education and Technology, 30(2), 239-254.\url{https://doi.org/DOI: 10.1007/s10956-020-09895-9 }

Mao, L., Liu, O. L., Roohr, K., Belur, V., Mulholland, M., Lee, H.-S., \& Pallant, A. (2018). Validation of automated scoring for a formative assessment that employs scientific argumentation. Educational Assessment, 23(2), 121-138. 

McGraw-Hill Education, C. (2014). Smarter balanced assessment consortium field test: Automated scoring research studies \url{https://www.smarterapp.org/documents/FieldTest_AutomatedScoringResearchStudies.pdf}

National Research Council. (2012). A framework for K-12 science education: Practices, crosscutting concepts, and core ideas. National Academies Press. 

Nehm, R. H., Ha, M., \& Mayfield, E. (2012). Transforming biology assessment with machine learning: automated scoring of written evolutionary explanations. Journal of Science Education and Technology, 21(1), 183-196.\url{ https://link-springer-com.proxy1.cl.msu.edu/content/pdf/10.1007%2Fs10956-011-9300-9.pdf}

Nehm, R. H., \& Haertig, H. (2012). Human vs. computer diagnosis of students’ natural selection knowledge: testing the efficacy of text analytic software. Journal of Science Education and Technology, 21(1), 56-73. \url{https://link-springer-com.proxy1.cl.msu.edu/content/pdf/10.1007%2Fs10956-011-9282-7.pdf}

NGSS Lead States. (2013). Next generation science standards: For states, by states. National Academies Press. 

Riordan, B., Bichler, S., Bradford, A., Chen, J. K., Wiley, K., Gerard, L., \& Linn, M. C. (2020). An empirical investigation of neural methods for content scoring of science explanations. Proceedings of the Fifteenth Workshop on Innovative Use of NLP for Building Educational Applications.

Urban-Lurain, M., Prevost, L., Haudek, K. C., Henry, E. N., Berry, M., \& Merrill, J. E. (2013). Using computerized lexical analysis of student writing to support just-in-time teaching in large enrollment STEM courses. 2013 IEEE Frontiers in education conference (FIE), 

Wang, C., Zhai, X., \& J., S. (In press). Applying Machine Learning to Assess Paper-Pencil Drawn Models of Optics. In X. Zhai \& J. Krajcik (Eds.), Uses of Artificial Intelligence in STEM Education. Oxford University Press. 

Williamson, D. M., Bennett, R. E., Lazer, S., Bernstein, J., Foltz, P. W., Landauer, T. K., \& Sweeney, K. (2010). Automated scoring for the assessment of common core standards. \url{http://professionals.collegeboard.com/ profdownload/Automated-Scoring-for-theAssessment-of- Common-Core-Standards.pdf}

Wilson, C., Haudek, K., Osborne, J., Stuhlsatz, M., Cheuk, T., Donovan, B., Bracey, Z., Mercado, M., \& Zhai, X. (2023). Using automated analysis to assess middle school students’ competence with scientific argumentation. Journal of Research in Science Teaching, 1-32. url{https://doi.org/10.1002/tea.21864}

Wu, X., He, X., Li, T., Liu, N., \& Zhai, X. (2023). Matching Exemplar as Next Sentence Prediction (MeNSP): Zero-shot Prompt Learning for Automatic Scoring in Science Education. In N. Wang \& e. al. (Eds.), AI in Education 2023 (Vol. LNAI 13916, pp. 1–13). Springer. \url{https://doi.org/https://doi.org/10.1007/978-3-031-36272-9_33}

Zhai, X. (2021). Practices and theories: How can machine learning assist in innovative assessment practices in science education. Journal of Science Education and Technology, 30(2), 139-149. \url{https://link.springer.com/article/10.1007/s10956-021-09901-8}

Zhai, X., Haudek, K., \& Ma, W. (2022). Assessing argumentation using machine learning and cognitive diagnostic modeling. Research in Science Education. \url{https://doi.org/https://doi.org/10.1007/s11165-022-10062-w}

Zhai, X., Haudek, K. C., Shi, L., Nehm, R., \& Urban-Lurain, M. (2020). From substitution to redefinition: A framework of machine learning-based science assessment. Journal of Research in Science Teaching, 57(9), 1430-1459. \url{https://doi.org/10.1002/tea.21658}

Zhai, X., Haudek, K. C., Stuhlsatz, M. A., \& Wilson, C. (2020). Evaluation of construct-irrelevant variance yielded by machine and human scoring of a science teacher PCK constructed response assessment. Studies in Educational Evaluation, 67, 100916. \url{https://doi.org/https://doi.org/10.1016/j.stueduc.2020.100916}

Zhai, X., He, P., \& Krajcik, J. (2022). Applying machine learning to automatically assess scientific models. Journal of Research in Science Teaching, 59(10), 1765-1794. 

Zhai, X., \& Krajcik, J. (2022). Pseudo AI Bias. arXiv preprint. \url{https://doi.org/10.48550/arXiv.2210.08141 }

Zhai, X., Krajcik, J., \& Pellegrino, J. (2021). On the validity of machine learning-based Next Generation Science Assessments: A validity inferential network. Journal of Science Education and Technology, 30(2), 298-312. \url{https://doi.org/10.1007/s10956-020-09879-9}

Zhai, X., \& Li, M. (2021). Validating a partial-credit scoring approach for multiple-choice science items: an application of fundamental ideas in science. International Journal of Science Education, 43(10), 1640-1666. \url{https://doi.org/https://doi.org/10.1080/09500693.2021.1923856}

Zhai, X., \& Nehm, R. (2023). AI and Formative Assessment: The Train Has Left the Station. Journal of Research in Science Teaching. \url{https://doi.org/DOI: 10.1002/tea.21885 }

Zhai, X., Neumann, K., \& Krajcik, J. (2023). AI for tackling STEM education challenges. Frontiers in Education, 8(1183030). \url{https://www.frontiersin.org/articles/10.3389/feduc.2023.1183030/full}

Zhai, X., Shi, L., \& Nehm, R. (2021). A meta-analysis of machine learning-based science assessments: Factors impacting machine-human score agreements. Journal of Science Education and Technology, 30(3), 361-379.\url{https://doi.org/10.1007/s10956-020-09875-z}

Zhai, X., Yin, Y., Pellegrino, J. W., Haudek, K. C., \& Shi, L. (2020). Applying machine learning in science assessment: a systematic review. Studies in Science Education, 56(1), 111-151. 

\end{hangparas}

\end{document}